# Embedded Model Control approach to robust control

Enrico Canuto, Wilber Acuna-Bravo, Andrés Molano-Jimenez, José Ospina, Carlos Perez-Montenegro

Politecnico di Torino, Dipartimento di Automatica e Informatica, Corso Duca degli Abruzzi 24, 10129 Torino, Italy
E-mail: enrico.canuto@polito.it

**Abstract:** Robust control design is mainly devoted to guarantee closed-loop stability of a model-based control law in presence of parametric and structural uncertainties. The control law is usually a complex feedback law which is derived from a (nonlinear) model, possibly complemented with some mathematical envelope of the model uncertainty. Stability may be guarantee with the help of some ignorance coefficients and restricting the feedback control effort with respect to the model-based design. Embedded Model Control shows that under certain conditions, the model-based control law must and can be kept intact under uncertainty, if the controllable dynamics is complemented by a suitable disturbance dynamics capable of real-time encoding the different uncertainties affecting the 'embedded model', i.e. the model which is both the design source and the core of the control unit. To be real-time updated the disturbance state is driven by an unpredictable input vector, called noise, which can be only estimated from the model error. The uncertainty (or plant)-based design concerns the noise estimator, as the model error may convey into the embedded model uncertainty components (parametric, cross-coupling, neglected dynamics) which are command-dependent and thus prone to destabilize the controlled plant. Separation of the components into the low and high frequency domain by the noise estimator allows to recover and guarantee stability, and to cancel the low frequency ones from the plant. Among the advantages, control algorithms are neatly and univocally related to the embedded model, the embedded model provides a real-time image of the plant, all control gains are tuned by fixing closed-loop eigenvalues. Last but not least, the resulting control unit has modular structure and algorithms, thus facilitating coding. A simulated case study helps to understand the key assets of the methodology.

**Key Words:** Embedded Model Control, disturbance, noise estimator, rejection, model error

## 1 INTRODUCTION

### 1.1 Goal of the paper and rationale

Robust control design [1] is mainly devoted to guarantee closed-loop stability of a model-based control law in presence of parametric uncertainties. The control law is usually a complex feedback law which is derived from a (nonlinear) model, possibly complemented with some mathematical envelope of the model uncertainty. Stability may be guarantee with the help of some ignorance coefficients and restricting the feedback control effort with respect to the model-based design. Embedded Model Control (EMC) [2] shows that under certain conditions, the model-based control law must and can be kept intact under uncertainty (a form of separation theorem), if the controllable dynamics is complemented by a suitable disturbance dynamics capable of real-time encoding the different uncertainties affecting the 'embedded model', i.e. the model which is both the design source and the core of the control unit. To be real-time updated the disturbance state is driven by an unpredictable input vector, called noise, which can be only estimated from the model error, defined as the difference between plant and model output. The uncertainty (or plant)-based design concerns the noise estimator, as the model error may convey into the embedded model uncertainty components (parametric, cross-coupling, neglected dynamics) which are command-dependent and thus prone to destabilize the controlled plant. Separation of the components into low and high frequency domains by the noise estimator allows recovering and guaranteeing stability, and rejecting the low frequency ones. Emphasis will be given to a single control unit, but extension to hierarchical and distributed control systems is embedded in the formulation. A case study will help to understand the key assets of the methodology.

### 1.2 Paper organization

A goal of the Embedded Model Control, only mentioned in the paper, is to offer a way for converting model/control architecture into the controller code, taking for granted that model/control architecture is not completely free, but constrained and guided by some basic statements (axioms and propositions when formulated). The surprising point is that interrogating literature with keywords like 'control' and 'axioms/propositions' only 'software' engineering papers are found [3], [4], [5], having null or weak relation to the immense bulk of control literature, as if control theory and implementation could run separate, leaving the latter to electrical and software engineering tools. Trying to fill in the gap, the paper is organized in a sequence of statements, aiming at fixing the basic and compulsory principles of model/control architecture, design and implementation. This is done with the help of the system and control theory at the foundations of the Embedded Model Control, with reference to control textbooks and papers.

Statements, that can be converted into formal propositions, are subdivided in two subsets: model and control. Model statements start from plant and model distinction and relation, leading to model error definition, as the key measurable variable of control design and performance (Section 2.2). To reduce unavoidable model error drifts, controllable dynamics must be enriched with a disturbance dynamics driven by noise (Section 2.3). As a key result, noise becomes the sole feedback channel (to be designed) from plant to model and control (Section 2.4). The resulting embedded model is further enriched (design model) with the set of command dependent discrepancies (parametric uncertainties, neglected dynamics) that are zeroed in the control unit, but are essential for robust design and assessment (Section 2.5). The design model may be implemented in the form of a numerical simulator, and as such it may surrogate the plant during design assessment and act as a test-bed for Monte Carlo trials.

Control statements start with the definition of the tracking error as a performance variable, and shows the standard control error (reference minus measure) is the sum of tracking and model error (Section 3.1). Since the former is model-based, the central theorem of the paper shows it can be brought to zero in the 'anti-causal' limit, and the latter can be approached 'in practice' by minimizing the tracking error, with no care of model discrepancies (model-based design, Section 3.2). Uncertainties are accommodated by the noise estimator, or better by the state predictor, which in the anti-causal limit becomes the sole responsible for plant stability (Section 3.3).

## 2 MODEL PRINCIPLES

### 2.1 Time, signals and the extended plant

Continuous time is denoted with $t$. Discrete-time instants are denoted with $t_i = iT$, assuming $T$ is the least and constant time unit of modeling and control. Real-valued discrete-time signals are indicated with $\mathbf{u}(i)$ and are assumed to be defined for $i \geq 0$. They distinguish from 'digital signals' $\tilde{\mathbf{u}}(i)$, which are defined as integer-valued and bounded. Computer-based control receives from and dispatch to plant only digital measures $\tilde{\mathbf{y}}(i)$ and commands $\tilde{\mathbf{u}}(i)$. $\tilde{\mathbf{y}}(i)$ is the sampled and digitized signal (ADC) of the sensor analogue output. $\tilde{\mathbf{u}}(i)$ is converted by a DAC into a step-wise analogue voltage driving the plant. To avoid treating integer values, $\tilde{\mathbf{u}}$ is digitized from a real-valued command $\mathbf{u}(i)$ (A/D conversion), whereas $\tilde{\mathbf{y}}(i)$ is converted to a real-valued measure $\mathbf{y}(i)$ (D/A conversion). The latter conversions take place in the control unit and should not be confused with conversions made by DAC and ADC. The 'extended plant' to be modeled and controlled is defined as the whole chain from $\mathbf{u}(i)$ to $\mathbf{y}(i)$, and is graphically represented by a 3D box as in Fig. 1. Discrete-time equations fit. The admissible set $\mathcal{U}$ of $\mathbf{u}(i)$ is defined here by a norm inequality

$$|\mathbf{u}(i)| \leq u_{\max}. \qquad (1)$$

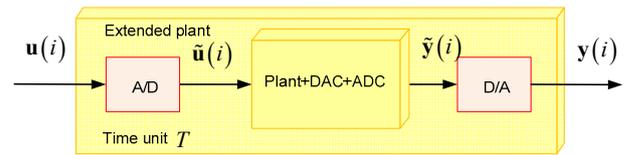

Fig. 1. Extended plant.

### 2.2 The fundamental input-output principles

*2.2.1 Statements*

Modern control design is a chapter of dynamical system theory [6], through concepts and methods like controllability, observability and feedback regulation [7]. 'Statements' should first lead to an appropriate model of the extended plant. Distinguishing between plant (reality) and (mathematical) model **M** leads to a pair of principles.

1) *Statement 1*. A model can run in parallel and synchronous (real-time) with the plant under the same admissible command **u** as in Fig. 2. The principle is seminal to subsequent formulation, as well as to control architecture, as it suggests that control units shall develop around the real-time model, henceforth indicated as the 'embedded model'. Restricting to computer-based control, a real-time model can only be discrete time and state variable [7], implying a time unit $T$ and a state $\mathbf{x}_c$ must be defined.

2) *Statement 2*. It concerns model performance, asserting that parallel plant and model can only be compared through the model error **e**, defined as the difference between the plant output **y** (measures) and the model output $\mathbf{y}_m$ as follows

$$\mathbf{e}(i) = \mathbf{y}(i) - \mathbf{y}_m(i). \qquad (2)$$

Error performance as a suitable norm $|\mathbf{e}|$ may be unbounded, because of unmeasurable, indescribable and uncertain discrepancies between reality (plant) and mathematics (model). Discrepancies may be due to command–independent actions (disturbance), parametric uncertainty, neglected interactions and dynamics, … The substantive attribute 'uncertainty' will replace 'discrepancy' throughout.

As a consequence of the second statement, model-based control design becomes unreliable in presence of an unbounded $|\mathbf{e}|$. A pair of alternative procedures are employed in the effort of overcoming the impasse: i) model error is ignored and the gains of a standard control law (for instance PID) are trial-and-error tuned either on the plant or on a surrogating simulator, ii) the model is complemented with input-output dynamical operators capable of enveloping the plant-to-model discrepancies. In the latter case $|\mathbf{e}|$ does not play an explicit role, but model and discrepancies are explicitly accounted for. The Embedded-Model-Control approach is rather different as a key effort is to devise a mechanism (see Sections 2.3 and 3.3) for guaranteeing that $|\mathbf{e}|$ is bounded for the class of admissible reference signals and uncertainty.

The mechanism consists of two parts:

1) modelling and real-time updating the past uncertainty (disturbance dynamics in Section 2.3 and noise estimator in Section 2.4), and cancelling it on the plant through the command **u** (Section 3.2),

2) protecting the embedded model from the high-frequency unmodelled uncertainty (Section 3.3) by confining the latter into the residual **e**.

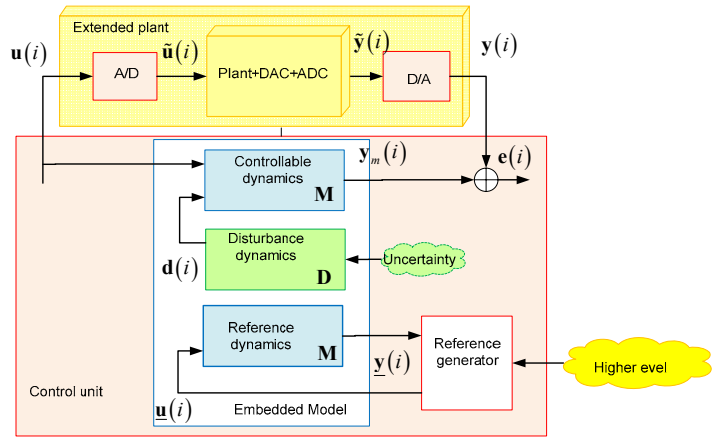

Fig. 2. The plant and the parallel embedded model as the core of the control unit.

The controllable dynamics **M** in Fig. 2 is defined as the causal relation between the command **u** and the model output $\mathbf{y}_m$, where the latter must be comparable to **y** so as to make use of **e** in (2). Assuming strict causality, linearity, time–invariance and a single time unit, the state equations of **M** are written as

$$\mathbf{x}_c(i+1) = A_c \mathbf{x}_c(i) + B_c \mathbf{u}(i) + \mathbf{d}(i),\ \mathbf{x}_c(0) = \mathbf{x}_{c0}$$
$$\mathbf{y}(i) = C_c \mathbf{x}_c(i) + \mathbf{e}(i) = \mathbf{y}_m(i) + \mathbf{e}(i)\quad (3)$$

where **d** (the disturbance vector), to be defined in Section 2.3, is the gate of the uncertainty as in Fig. 2

The controllable dynamics is usually accompanied by the reference dynamics expressing reference state, output and command trajectories, that are denoted with $\underline{\mathbf{x}}$, $\underline{\mathbf{y}}$ and $\underline{\mathbf{u}}$ respectively. The open-loop command $\underline{\mathbf{u}}$ is generated by a reference generator (see Fig. 2) in charge of matching requests coming from a higher level unit (supervisor, operator, mission program) with the reference output $\underline{\mathbf{y}}$. Under linear assumption, the reference dynamics repeats (3) as follows

$$\underline{\mathbf{x}}(i+1) = A_c \underline{\mathbf{x}}(i) + B_c \underline{\mathbf{u}}(i) + \underline{\mathbf{d}}(i),\ \underline{\mathbf{x}}(0) = \underline{\mathbf{x}}_0$$
$$\underline{\mathbf{y}}(i) = C_c \underline{\mathbf{x}}(i) \quad (4)$$

where $\underline{\mathbf{d}}(i)$ is a priori known, but set to zero in this paper. $\underline{\mathbf{u}}$ is assumed to bounded as in (1).

*2.2.2 Case study*

Consider the attitude of a satellite as in [8] and [9], where the attitude sensor is mounted on a flexible axial appendage. The sensor is affected by bias (0.1 mrad) and random errors (0.5 mrad, 1σ). A gyro affected by drift, bias and random error (0.1 mrad/s, 1σ) is mounted on the spacecraft. Restricting to a single degree-of-freedom (DoF) as in [9], the transfer function of the design model (see 2.5) from the command torque $u$ to the measured attitude $q$ is fourth order as follows

$$P(s) = \frac{q}{u}(s) = \frac{1}{J_0(1+\partial J)s(s+1/\tau)}\frac{1}{v^2 + 2\zeta_f v + 1},\ (5)$$

and has the parameter values

$$v = s/\omega_f,\ \omega_f \geq 6\ \text{rad/s},\ \zeta_f \geq 0.002,\ \tau \leq 60\ \text{s},\quad (6)$$

In (5) $J_0 = 1200\ \text{kgm}^2$ is the total inertia ($\partial J = \pm 0.2$), $\omega_f$ and $\zeta_f$ are the lowest angular frequency and damping of the flexible link between spacecraft and star tracker, $\tau$ is the inertia/friction time constant. The gyro measure $y_g(i)$ is sampled at $1/T = 100\ \text{Hz}$. The attitude measure $y_q(i_k)$ is sampled at $1/T_q$, $T_q = N_q T$, with $N_q = 10$, and $i_k = kN_q$. The attitude $q$ is the inertial rotation around the spacecraft axis forced to track a variable reference $\bar{q}$. The angular rate, in angular units, measured by the gyro, is denoted with $\omega$. The command acceleration (dispatched to reaction wheels) is $a_u = T^2 u/J$ and is bounded by $|a_u/T^2| \leq 25\ \text{mrad/s}^2$. Slew rate limit of the order of $|\dot{a}_u T^2| \leq 0.250\ \text{rad/s}^3$ is accounted for by the reference generator.

The transfer function (5) is typical of single DoF rotation and translation devices when position sensor is not collocated. Table 1 shows root mean square (RMS), bias and drift of the measurement errors, and the noise ratio between gyro and attitude sensor. Noise RMS accounts for quantization errors.

Table 1 Measurement errors.

| No | Error | Unit | RMS value | Comment |
|---|---|---|---|---|
| 0 | Attitude noise | mrad | 0.5 | |
| 1 | Attitude bias | mrad | 0.1 | |
| 2 | Gyro noise | mrad/s | 0.1 | |
| 3 | Gyro bias | mrad/s | 0.01 | |
| 4 | Gyro drift noise | mrad/s | 0.001 | |
| 5 | Noise ratio gyro/attitude | mrad/mrad | 0.0064 | @ 10 Hz |

The embedded model **M** only accounts for rigid motion, confining flexible dynamics into model error. The total of the disturbance accelerations (drag, gravity gradient, friction, ...) is denoted in (7) by $d_q$. State and multi-rate output equations in (3) can be obtained from the following vectors and matrices

$$\mathbf{x}_c = \begin{bmatrix} q_d \\ \omega_d \end{bmatrix}, A_c = \begin{bmatrix} 1 & 1 \\ 0 & 1 \end{bmatrix}, B_c = \begin{bmatrix} 1/2 \\ 1 \end{bmatrix}$$

$$\mathbf{y} = \begin{bmatrix} y_q \\ y_g \end{bmatrix}, C_c = \begin{bmatrix} 1 & 0 \\ 0 & 1 \end{bmatrix}, \mathbf{e} = \begin{bmatrix} e_q \\ e_g \end{bmatrix}, \mathbf{d} = \begin{bmatrix} d_q \\ d_g \end{bmatrix}, \quad (7)$$

$$y_q(i) = \begin{cases} y_q(i_k = kN_a) \\ \text{NA}, i \neq kN_a = 10k \end{cases}$$

where $q_d$ and $\omega_d$ are 'dirty' variables to include systematic errors as in [9]. The dirty rate $\omega_d$ is corrected by a disturbance $d_g$ so as to recover the 'true' rate driving $q_d$.

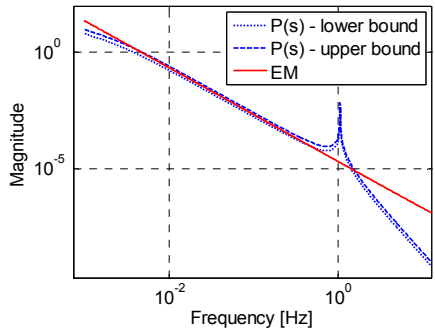

Fig. 3. Magnitude of the Bode diagrams.

Fig. 3 shows the Bode plots of the uncertain design model (5) (upper and lower bounds) and of the embedded model. Equation (7) is observable and controllable. According to (4) the reference attitude satisfies the same dynamics as in (3) and (7), but free of disturbance, i.e.

$$\underline{\mathbf{x}}(i+1) = A_c\underline{\mathbf{x}}(i) + B_c\underline{u}(i), \underline{\mathbf{x}}(0) = \underline{\mathbf{x}}_0$$
$$\underline{\mathbf{y}}(i) = C_c\underline{\mathbf{x}}(i), \underline{\mathbf{x}}^T = \begin{bmatrix} \underline{q} & \underline{\omega} \end{bmatrix} , \quad (8)$$

where $\underline{u}$ is the open-loop command.

## 2.3 The uncertainty principles

### 2.3.1 Statements

Feedback regulation implicitly decreases output sensitivity to discrepancies. Sensitivity may be further abated by explicitly rejecting disturbance as in [10], [11], [12], which leads to disturbance modeling. A pair of statements are of guideline.

1) *Statement 3*. Model error **e** is the sole available measure of the uncertain discrepancies ('uncertainty'), since it encodes the current outcome of the past discrepancies.
2) *Statement 4*. Model error can be elaborated and accumulated in a state vector $\mathbf{x}_d$ (disturbance state), ready to correct $\mathbf{x}_c$. Formally, an observable input-output dynamics **D** as in [10], [11], [12] and Fig. 2 must be built, from an input **w** to an output **d**, the latter forcing **M** in parallel to **u**. As a result, $\mathbf{x}_d$ encodes the past accumulated discrepancies, whereas **w** encodes the past and future independent uncertainty (innovation) capable of updating $\mathbf{x}_d$. Independence of future derives from causality, whereas independence of past answers the principle of not delaying disturbance updating. For such reasons **w**, called 'noise', should be treated as an element of a class $\mathbb{W}$ of arbitrary and bounded zero-mean signals, and statistically as a bounded-variance discrete-time white noise. In other terms, no state equation exists relating past to future of **w**. The class $\mathbb{W}$ is a set of signals closed under concatenation, thus favouring arbitrariness in the time domain, and flat spectrum in the frequency domain.

Disturbance dynamics is widely treated in the literature [10], [11], [12], [13]; the opposite can be said of the noise design [2], [8]. As a conclusion, the embedded model in Fig. 2 is forced by two input vectors: $\mathbf{u}(i)$ is known since it is computed at any step $i$ by the control unit, $\mathbf{w}(i)$ is defined to be unknown and unpredictable. How to retrieve **w** at any step $i$ is the subject of Section 2.4.

### 2.3.2 State equations

Assuming linearity, the disturbance dynamics is written as

$$\mathbf{x}_d(i+1) = A_d\mathbf{x}_d(i) + G_d\mathbf{w}(i), \mathbf{x}_d(0) = \mathbf{x}_{d0}$$
$$\mathbf{d}(i) = H_c\mathbf{x}_d(i) + G_c\mathbf{w}(i) + \mathbf{h}(\mathbf{x}_c) \quad , \quad (9)$$

where **h** may be referred to as cross-coupling term, and is discussed in Section 2.5. The state $\mathbf{x}_d$ must be observable from the model output $\mathbf{y}_m$. The free response of (9) is used to model 'deterministic' components as in the Internal Model Control [14], assuming $\mathbf{x}_{d0}$ is known. Usually $(A_d, G_d)$ is controllable. Otherwise, $\mathbf{x}_d$ splits into controllable and non controllable entries; since the latter ones only account for deterministic components, they do not need to be observable. When $H_c = B_cM_c$, $G_c = B_cN_c$, $\mathbf{h} = B_c\mathbf{m}$, **d** is called collocated since it may add to the command. The paper treats the generic case, except for the nonlinear term **h** which is assumed to be collocated.

The set of the state equations (3), (4) and (9) constitute the a priori embedded model, where 'a priori' marks the fact that the model signals because of the command and noise classes $\mathcal{U}$ and $\mathbb{W}$ constitute a class until a realization of $\mathbf{w}(i)$ becomes known. Properties of the embedded-model signal classes can be estimated either analytically or by running Monte Carlo trials on a suitable simulator.

### 2.3.3 Case study

The disturbance vector **d** in (9) splits into two components: (i) the acceleration disturbance $d_u = a + w_u$, and (ii) the velocity disturbance $d_g = a/2 + s_g$, where $s_g$ compensates the gyro systematic errors of $\omega_d$. A 2$^{nd}$ order dynamics describes $d_u$, and a 1$^{st}$ order one pertains to $d_g$, as follows

$$\mathbf{x}_d = \begin{bmatrix} s_g \\ a \\ s \end{bmatrix}, A_d = \begin{bmatrix} 1 & 0 & 0 \\ 0 & 1 & 1 \\ 0 & 0 & 1 \end{bmatrix}, G_d = \begin{bmatrix} 1 & 0 & 0 & 0 \\ 0 & 0 & 1 & 0 \\ 0 & 0 & 0 & 1 \end{bmatrix}$$

$$\mathbf{w} = \begin{bmatrix} w_g \\ w_u \\ w_a \\ w_s \end{bmatrix}, G_c = \begin{bmatrix} 0 & 0 & 0 & 0 \\ 0 & 1 & 0 & 0 \end{bmatrix}, H_d = \begin{bmatrix} 1 & 1/2 & 0 \\ 0 & 1 & 0 \end{bmatrix} \quad . \quad (10)$$

Noise entries are statistically independent. Equations (7) and (10) are observable, $\mathbf{d}$ is not collocated because of $d_g$.

The block-diagram of (7), (8) and (10) is shown in Fig. 4. Boxes marked with $\Sigma$ denote discrete-time integrators.

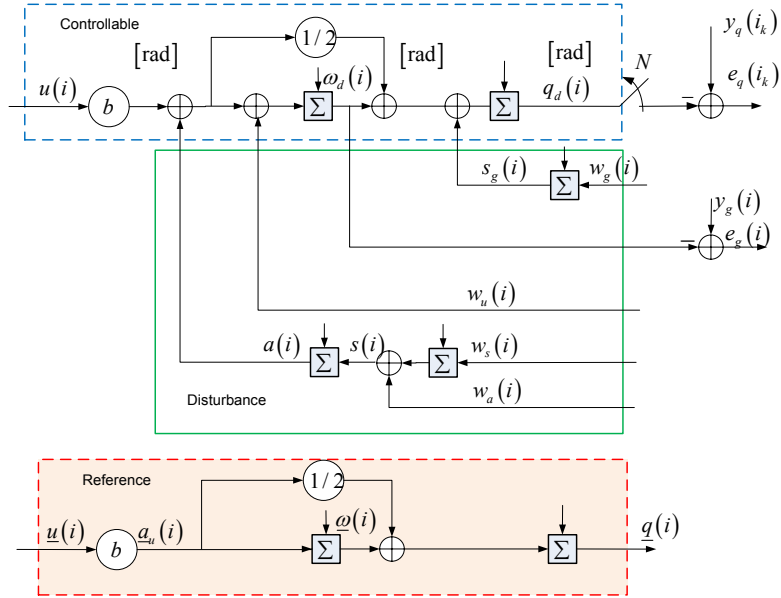

Fig. 4. Block-diagram of the embedded model for case study.

### 2.4 The central principles: noise estimation

*2.4.1 Statements*

Statements repeat Kalman filter results, but from a generic standpoint.

1) *Statement 5.* Two alternative mechanisms can generate noise: i) pseudo-random extraction, ii) 'estimation' from a correlated realization. The former would respect noise statistical properties, the latter, to be adopted, reveals the residual discrepancies that are hidden in the model error to the benefit of the embedded model, as it can be driven to approach the plant and to bound $|\mathbf{e}|$. Complexity and uncertainty of discrepancies may suggest to abandon the statistical framework in favor of a bounded arbitrariness, which entails command independence. The latter assumption, referred to as 'Kalman assumption' in Section 2.5, does not hold in general, but suggests the following noise estimation algorithm.

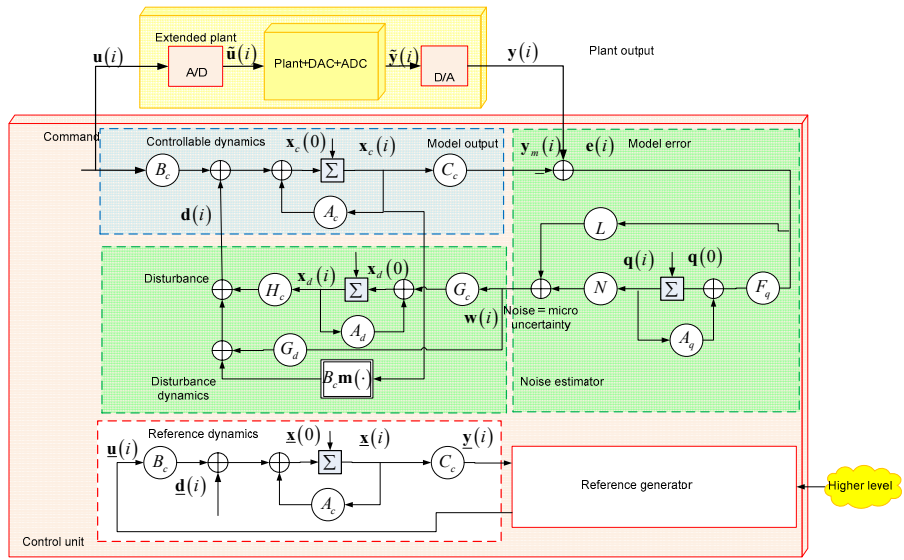

Fig. 5. Block-diagram of the embedded model plus noise estimator.

2) *Statement 6.* Under Kalman assumption and the equations (3) and (9), the 'noise estimator'[2], [8] is a linear dynamic system as follows

$$\overline{\mathbf{w}}(i) = L\overline{\mathbf{e}}(i) + N\mathbf{q}(i)$$
$$\mathbf{q}(i+1) = A_q \mathbf{q}(i) + B_q C_c \overline{\mathbf{e}}(i), \mathbf{q}(0) = \mathbf{q}_0 \quad (11)$$

where matrices $L$, $N$, $A_q$ and $B_q$ must be designed to make the closed-loop system (3), (9) and (11) asymptotically stable and minimum variance. The closed-loop system is a one-step state predictor, as it provides, at any time $i$, predictions $\mathbf{x}_c(i+1)$ and $\mathbf{x}_d(i+1)$. To mark that state variables and their predictions, when updated by the noise realization (11) become realizations of the embedded-model signals, the 'hat' notation as in $\hat{\mathbf{x}}_c(i)$ is applied to the state variables and to the model output in (3), (9) and (11) in analogy with prediction theory [15]. In fact, $\hat{\mathbf{x}}_c(i)$ is a one-step prediction $\hat{\mathbf{x}}_c(i/i-1)$ based on the past sequence $\{\mathbf{y}(i-k), k>0\}$ of the plant output. Noise estimation and model error are marked with a 'bar' notation as in (11) to mean dependence on the current output $\mathbf{y}(i)$.

According to statement 6, the embedded model implemented by the control unit can be kept as a real-time realization of the 'a priori' embedded model consisting of (3), (4) and (9). To this end the attribute 'a posteriori' applies. Fig. 5 shows the ensemble of embedded model and noise estimator.

A corollary that the authors never met in the literature, yet is a cornerstone of Embedded Model Control architecture and design is the following.

*Corollary 1*. The a posteriori noise $\bar{\mathbf{w}}$ is the sole and unique feedback channel from plant to model and control, thus becoming responsible for stability and performance.

*Remark 1*. Noise estimators in Kalman filtering are static, because noise is implicitly forced to affect all state variables, different from here, where noise design is a modelling stage as in [8]. Thus, when the noise components of the embedded model are insufficient to guarantee stability by means of a proportional feedback, a dynamic feedback must be employed.

*Remark 2*. Noise estimator concept facilitates multi-rate treatment as in Section 2.4.2 (see also [9]), since noise is estimated only when the relevant model error becomes available.

### 2.4.2 Case study

The noise estimators of (7) and (10) must be multi-rate and dynamic. A decoupling design is adopted as in [9] exploiting the different noise RMS and sampling rate of the measurements (see Table 1).. Specifically, $w_g$ driving $s_g$ in (10) is estimated from the attitude error $\bar{e}_q = y_q - \hat{q}$, whereas the remaining noise entries in (10) from $\bar{e}_g = y_g - \hat{\omega}$. Since $w_g$ drives a pair of state variables, $\hat{s}_g$ and $\hat{q}$, a dynamic feedback is necessary to guarantee closed-loop stability. The following noise estimator results

$$\bar{w}_g(i_k) = l_q \bar{e}_q(i_k) + m_q p(i_k)$$
$$p(i_k + N_q) = (1-\beta_q) p(i_k) + \bar{e}_q(i_k). \quad (12)$$
$$\begin{bmatrix} \bar{w}_u & \bar{w}_a & \bar{w}_s \end{bmatrix}^T (i) = L\bar{e}_g(i)$$

A proof of decoupling may proceed along two steps: (i) the low noise ratio in Table 1, last row, forces to zero the gains of $\bar{e}_q$, that should add to $L\bar{e}_g$ in the last row of (12), (ii) if the time constant of the state predictor driven by $\bar{e}_g$ could be made shorter than the attitude time unit $T_a$, the prediction error would become uncorrelated during $T_a$, thus making the attitude error correction useless.

It is of interest to compare the upper part of (12) with a static observer feedback to be employed by Kalman filters, namely

$$\begin{bmatrix} \bar{w}_q \\ \bar{w}_g \end{bmatrix} (i_k) = L_q \bar{e}_q(i_k), \quad (13)$$

which implies $d_g$ in (10) must be corrected into

$$d_g(i) = a(i)/2 + s_g(i) + w_q(i). \quad (14)$$

The correction amounts to add the 'parasitic' noise $w_q$ to $\omega_d$, in contrast with the smoothness of the systematic errors affecting $\omega_d$.

### 2.5 The central issue: the uncertainty model

Noise arbitrariness, entailing command independence, cannot capture the model error complexity. For instance, parametric uncertainty, neglected cross-coupling and dynamics are command-driven, and as such they contrast noise assumptions. The solution, well known in the literature, is to build an appropriate model of the uncertainty capable of detailing the different components: see for instance the uncertainty models built by means of the Linear Fractional Transformation [1]. A similar approach is adopted here, but an intermediate model, called 'design model', is created, which is the combination of the embedded and uncertainty models. The embedded model alone must be implemented in the control unit, setting all the uncertainty to zero except the disturbance dynamics in (9), and the known part of $\mathbf{m}(\cdot)$ in $\mathbf{h}(\cdot)$, if any. An alternative is pursued by adaptive control and estimation schemes [16], [17], where command/state dependence is accounted for by multiplying noise components times the same command/state signals.

*Statement 7*. Command/state dependent uncertainty is modeled in two ways: (i) the cross-coupling $\mathbf{h}(\mathbf{x}_c)$ in (9) accounts for parametric uncertainty, including nonlinear approximations, (ii) the model error $\mathbf{e}$ accounts for the neglected dynamics (unstructured uncertainty) as follows

$$\mathbf{e}(i) = \mathbf{E}(\mathbf{y}_m) + \mathbf{w}_y(i), \quad (15)$$

where superposition has been assumed of $\mathbf{w}_y$ (the measurement noise) and of the (uncertain) fractional error dynamics $\mathbf{E}$ mapping the model output $\mathbf{y}_m$ into $\mathbf{e}$. The proof follows by rewriting (15) as

$$\mathbf{e}(i) = \mathbf{P}(\mathbf{M}^{-1}\mathbf{y}_m) - \mathbf{y}_m + \mathbf{w}_y(i), \quad (16)$$

where the former term in the right-hand side exists if $\dim \mathbf{y}_m = \dim \mathbf{u}$, and the response of $\mathbf{P}$ to a step $\mathbf{u}_0$ is delayed and smaller than $\mathbf{M}$, i.e.

$$\lim_{t \to \varepsilon > 0} |\mathbf{P}(\mathbf{u}_0)|/|\mathbf{M}(\mathbf{u}_0)| \leq o(\varepsilon), \quad (17)$$

assuming $\varepsilon$ arbitrarily small. In other terms, the short-term transient of the design model, emulating the plant, is of higher order than the model. If $\mathbf{P}$ and $\mathbf{M}$ are rational transfer functions, (17) converts into

$$\lim_{f \to \infty} |\mathbf{M}^{-1}(f)\mathbf{P}(f) - I| = \lim_{f \to \infty} |\mathbf{E}(f)| \to 1. \quad (18)$$

Zero fractional dynamics $\mathbf{E} \equiv 0$ and zero cross-coupling $\mathbf{m} = 0$, are referred to as 'Kalman assumption', since they are necessary for Kalman filter to be unbiased and efficient.

*2.5.1 Case study*

The embedded model (7), written in Laplace transform, as

$$\mathbf{M}(s) = 1/(J_0 s^2), \quad (19)$$

and the fractional error dynamics

$$\mathbf{E}(s) \cong -\frac{\frac{1-\partial J}{s\tau+1} + \partial J + (v^2 + 2\zeta v)}{v^2 + 2\zeta v + 1} \quad (20)$$

satisfy (18). Parametric uncertainty because of $\partial J$ and of the friction time constant $\tau$ enter (20) as low-frequency contributions. $\tau$ is completely ignored by the controllable dynamics (7).

## 3 CONTROL PRINCIPLES

Embedded and design models, as well as noise estimators are ingredients to control architecture and design, which latter splits into two stages: (i) the model-based design concerns the control law in charge of the command $\mathbf{u}$ and employs real-time data from the a posteriori embedded model and the reference signals, (ii) the uncertainty-based design is in charge of tuning the estimator noise gains so as to guarantee stability in the presence of the unmodelled uncertainty. Preliminary is the definition of the control performance errors.

### 3.1 Performance statements

*Statement 8*. Performance is expressed through the 'a priori' tracking error

$$\begin{aligned}\underline{\mathbf{e}}(i) &= \underline{\mathbf{x}}(i) - \mathbf{x}_c(i) - Q\mathbf{x}_d(i) \\ \underline{\mathbf{e}}_y(i) &= \underline{\mathbf{y}}(i) - \mathbf{y}_m(i) = C_c \underline{\mathbf{e}}(i)\end{aligned} \quad (21)$$

where $\underline{\mathbf{x}}$ and $\underline{\mathbf{y}}$ are reference state and output satisfying the reference dynamics (8), and the matrix $Q$ entering the control law (30) together with $M_c$, is the solution of the Sylvester equation [2]

$$\begin{bmatrix} H_c + QA_d \\ 0 \end{bmatrix} = \begin{bmatrix} A_c & B_c \\ C_c & 0 \end{bmatrix} \begin{bmatrix} Q \\ M_c \end{bmatrix}. \quad (22)$$

Matrices $Q$ and $M_c$ only depend on the embedded model matrices in (3) and in (9). The tracking error $\underline{\mathbf{e}}$ is available either mathematically or through simulation, since only the a posteriori state and output variables, $\hat{\mathbf{x}}_c$, $\hat{\mathbf{x}}_d$ and $\hat{\mathbf{y}}_m$ are available in the control unit.

*Statement 9*. The only measurable errors are: (i) the a posteriori tracking error

$$\hat{\underline{\mathbf{e}}}(i) = \underline{\mathbf{x}}(i) - \hat{\mathbf{x}}_c(i) - Q\hat{\mathbf{x}}_d(i) = \underline{\mathbf{e}}(i) + \hat{\mathbf{e}}_c(i), \quad (23)$$

which is the sum of the a priori tracking error $\underline{\mathbf{e}}$ and of the prediction error defined by

$$\hat{\mathbf{e}}_c = \mathbf{x}_c(i) - \hat{\mathbf{x}}_c(i/i-1), \quad (24)$$

(ii) the a posteriori model error

$$\overline{\mathbf{e}}(i) = \mathbf{y}(i) - \hat{\mathbf{y}}_m(i) = C_c \hat{\mathbf{e}}_c(i) + \mathbf{e}(i). \quad (25)$$

*Remark 3*. Decomposition (23) and (25) of a posteriori errors into 'a priori' ones is justified by any realization of a priori errors being only known as possible element of a signal class.

The following corollary links errors in (23) and (25) to the standard control error defined as the difference between reference and plant output.

*Corollary 2*. The control error (jitter) $\mathbf{e}_y = \underline{\mathbf{y}} - \mathbf{y}$ is the sum of the a posteriori tracking and model errors as follows

$$\mathbf{e}_y(i) = \underline{\mathbf{y}}(i) - \mathbf{y}(i) = \hat{\underline{\mathbf{e}}}_y(i) - \overline{\mathbf{e}}(i). \quad (26)$$

Model- and uncertainty-based designs emerge from (26): (i) the model-based aims to make $\hat{\underline{\mathbf{e}}}_y(i)$ 'negligible', which is feasible since reference and prediction are model variables; (ii) the uncertainty-based aims to make $\overline{\mathbf{e}}(i)$ bounded notwithstanding uncertainty and discrepancies. Were those aims achieved, (26) would simplify to $\mathbf{e}_y(i) \cong -\overline{\mathbf{e}}(i)$, implying the jitter to be the opposite of the model error. The actual performance is expressed through some error norm. Errors are kept as negligible when their norm is below the relevant quantization error.

The a posteriori model error in the form

$$\overline{\mathbf{e}}(t/t) = \mathbf{y}(t) - \hat{\mathbf{y}}_m(t/t-\tau), \quad (27)$$

is employed by prediction and identification theory [15] under the name of prediction error, a name here reserved to (23) which is fully model-based. The model error

$$e = y(i) - y_m(i), \quad (28)$$

is employed as a tracking (or output) error in the Model Reference Adaptive Control [16], where the model acts as a reference to be tracked by the output. A similar approach and nomenclature (reference error) is adopted by the Internal Model Principle [14], and by the Model Predictive Control scheme [18], in which case the model role is played by the reference signal. On the contrary, Embedded Model Control fully distinguishes between model and control errors as pointed out by Corollary 2.

*3.1.1 Case study*

Target and simulated performance are expressed in Table 2 in terms of the peak absolute value (max), of the root mean square and of the mean value of the a priori tracking errors. Simulated results refer to Fig. 6, Fig. 8, Fig. 9 and Fig. 12. The attitude mean value is very close to the sensor bias of Table 1 (0.1 mrad), as the latter cannot be estimated and rejected. The rate mean value is instead less than the gyro bias (0.01 mrad/s), as the latter has been estimated and rejected. Both attitude and rate RMS are lower than sensor noise as a consequence of the state predictor bandwidth in Table 3.

Table 2 Target and simulated performance.

| No | Tracking error | Target | | Simulated | | |
|---|---|---|---|---|---|---|
| | | Max | RMS | Max | RMS | Mean |
| 0 | Attitude [mrad] | 0.6 | 0.2 | 0.5 | 0.15 | 0.12 |
| 1 | Rate [mrad/s] | 0.25 | 0.18 | 0.07 | 0.06 | <0.002 |

### 3.2 Model-based control theorem

*3.2.1 Fundamental theorem*

When and how the a posteriori tracking errors $\hat{\underline{e}}$ and $\hat{\underline{e}}_y = C_c \hat{\underline{e}}$ in (26) can be made negligible is expressed by the following theorem.

*Theorem 1.* The mean value of the a posteriori tracking error $\hat{\underline{e}}$ can be brought asymptotically to zero; the error norm $|\hat{\underline{e}}|$ can only be bounded because of causality.

*Proof.* The proof may be obtained from the following error equation

$$\hat{\underline{e}}(i+1) = (A_c - B_c K)\hat{\underline{e}}(i) - (G_c + QG_d)\overline{\mathbf{w}}(i)$$
$$\hat{\underline{e}}(0) = \hat{\underline{e}}_0$$
$$\overline{\mathbf{w}}(i) = N\mathbf{q}(i) + L\overline{\mathbf{e}}(i)$$
$$\mathbf{q}(i+1) = A_q \mathbf{q}(i) + B_q C_c \overline{\mathbf{e}}(i), \quad \mathbf{q}(0) = \mathbf{q}_0$$

, (29)

if and only if $A_c - B_c K$ is asymptotically stable, and $\overline{\mathbf{w}}$ is zero-mean and bounded.

Equation (29) is proved assuming the following control law

$$\mathbf{u}(i) = \underline{\mathbf{u}}(i) + K\hat{\underline{e}}(i) - \hat{\mathbf{d}}_u(i)$$
$$\hat{\mathbf{d}}_u = M_c \hat{\mathbf{x}}_d + \underline{\mathbf{m}}(\hat{\mathbf{x}}_c)$$

, (30)

where $\underline{\mathbf{m}}$ is the known model of $\mathbf{m}$, and the estimated noise $\overline{\mathbf{w}}(i)$ cannot enter (30), being simultaneous to $\mathbf{u}(i)$ (the causality constraint). In other terms, all the components of (30) must be one-step predictions, but the noise prediction is zero. The feedback matrix $K$ must be designed to make $A_c - B_c K$ asymptotically stable, which is feasible since $(A_c, B_c)$ is controllable.

The next step is to define

1) the output disturbance $\mathbf{d}_y$ as follows
$$\mathbf{x}_y(i+1) = A_c \mathbf{x}_y(i) + \mathbf{d}(i) - B_c \underline{\mathbf{m}}(\hat{\mathbf{x}}_c), \quad \mathbf{x}_y(0) = 0$$
$$\mathbf{d}_y(i) = C_c \mathbf{x}_y(i)$$
, (31)

2) the corrected prediction error
$$\hat{\boldsymbol{\eta}}_c(i) = \hat{\mathbf{e}}_c(i) - \mathbf{x}_y(i), \quad (32)$$

3) and the corrected tracking error
$$\underline{\boldsymbol{\eta}}(i) = \underline{\mathbf{e}}(i) + Q\hat{\mathbf{e}}_d(i) + \mathbf{x}_y(i), \quad (33)$$

where the prediction error $\hat{\mathbf{e}}_d = \mathbf{x}_d - \hat{\mathbf{x}}_d$ has been employed. The sum of the errors in (32) and (33) provides the a posteriori tracking error of the equation (29), i.e.

$$\underline{\boldsymbol{\eta}}(i) + \hat{\boldsymbol{\eta}}_c(i) = \underline{\mathbf{e}}(i) + \hat{\mathbf{e}}_c(i) = \hat{\underline{\mathbf{e}}}(i), \quad (34)$$

which implies (29) can be obtained by adding state equations of (32) and (33). Using Sylvester equation (22), adding the prediction error equation

$$\hat{\boldsymbol{\eta}}_c(i+1) = A_c \hat{\boldsymbol{\eta}}_c(i) - H_c \hat{\mathbf{x}}_d(i) - G_c \overline{\mathbf{w}}(i), \quad (35)$$

and the tracking error equation

$$\underline{\boldsymbol{\eta}}(i+1) = A_c \underline{\boldsymbol{\eta}}(i) - B_c K \hat{\underline{\mathbf{e}}}(i) - QG_d \overline{\mathbf{w}}(i) +$$
$$+ (-QA_d + A_c Q + BM_c)\hat{\mathbf{x}}_d(i)$$
, (36)

equation (29) follows.

The properties of $\overline{\mathbf{w}}$ are guaranteed (i) by the prediction error equation (35) being asymptotically stable, and (ii) by $\mathbf{e}$ and the cross-coupling error

$$\Delta \mathbf{m}(\cdot) = \mathbf{m}(\mathbf{x}_c) - \underline{\mathbf{m}}(\hat{\mathbf{x}}_c) \quad (37)$$

being bounded. The latter condition is immediately satisfied under Kalman assumption, since $\mathbf{e} = \mathbf{w}_y$, and $\Delta \mathbf{m} = 0$.

The following corollary states that (29) cannot be improved.

*Corollary 3.* Noise rejection can only enter (30) if delayed, i.e. through $\overline{\mathbf{w}}(i-k)$, $k > 0$, which implies (29) becomes forced by the difference $\overline{\mathbf{w}}(i) - \overline{\mathbf{w}}(i-k) \neq 0$, and the latter has a norm larger than $\overline{\mathbf{w}}(i)$ under sample independence. The ideal, unrealizable case

$$\hat{\underline{\mathbf{e}}}(i+1) = (A_c - B_c K)\hat{\underline{\mathbf{e}}}(i) \Rightarrow \lim_{i \to \infty} \hat{\underline{\mathbf{e}}}(i) = 0 \quad (38)$$

is referred to as the 'anti-causal' limit. The asymptote (38) can be approached by moving the discrete-time eigenvalues of $A_c - B_c K$ toward zero. Such a design guideline agrees with the model-based control law (30), but contrasts standard robust design [1], [19], [20], where feedback gains are responsible for uncertainty-based stability and performance. In the anti-causal limit, prediction and tracking errors asymptotically substitute each other, which is assumed hereafter, since $\underline{\mathbf{e}}(i) + \hat{\mathbf{e}}_c(i) = \hat{\underline{\mathbf{e}}}(i)$ from (23).

*Statement 10.* The gain $K$ in (29) must be designed to minimize the effect of the not rejected noise components on the tracking error, thus approaching the anti-causal limit. The only limit comes from bounds on the control authority (range and slew rate), as in the linear quadratic optimal control. In this sense, the feedback control design only depends on the causal uncertainty expressed by the noise, and not on parametric and unstructured uncertainty (model-based design). A control design approaching the anti-causal limit is referred to as 'standard'.

*Remark 4.* There is no prohibition of designing $K$ to withstand uncertainty, except that separation of model- and uncertainty-based design is lost.

*3.2.2 Case study*

The control law following (30) becomes

$$a_u(i) = \underline{a}_u + k_q(q - \hat{q}_d) + k_\omega(\omega - \hat{\omega}_d - s_g) - \hat{a}, \quad (39)$$

where the rate tracking error $\omega - \hat{\omega}_d$ has been corrected with $s_g$ to compensate for gyro systematic errors.

The complete block-diagram of the control unit around the embedded model of Fig. 4 is in Fig. 7. The box marked D is a unit delay.

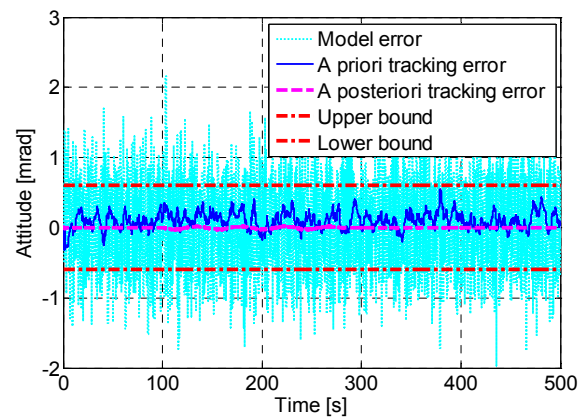

Fig. 6. True and estimated tracking error, and model error.

Fig. 6 shows a realization of the a priori (solid) and a posteriori tracking (dashed) errors of the attitude $q$ in (5), and compares them with the estimated model error (dotted). The a posteriori tracking error is practically zero - marking standard design -, whereas the a priori tracking error is the opposite of the model error (not perceivable from Fig. 6), but is free of the high frequency sensor noise, though it is affected by bias.

Were the noise estimator implemented as a static feedback as in (13), the tracking error would become noisier owing to the 'parasitic' noise $w_q$ in (14). Fig. 8 and Fig. 9 allow comparison between the dynamic design in (12) and the static one in (13). Enlargement of the 'static' error in Fig. 9 reveals significant oscillations of the flexible link passing through $\overline{w}_q$. A remedy would be either a narrower bandwidth or a notch filter, which latter artifice, suffering of tuning, is not in the need of the dynamic design.

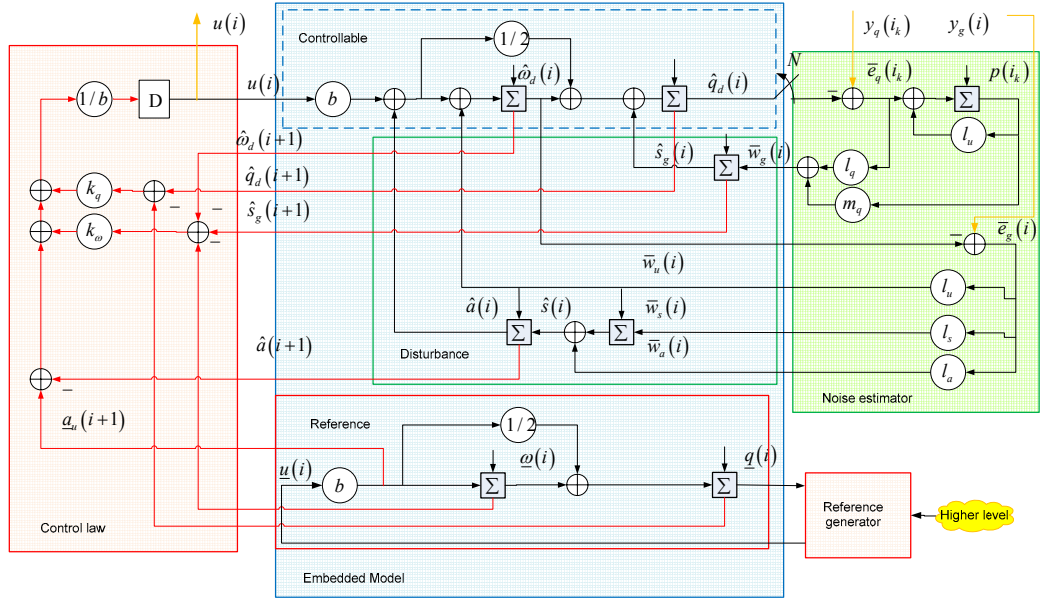

Fig. 7. Block-diagram of the control unit around the embedded model.

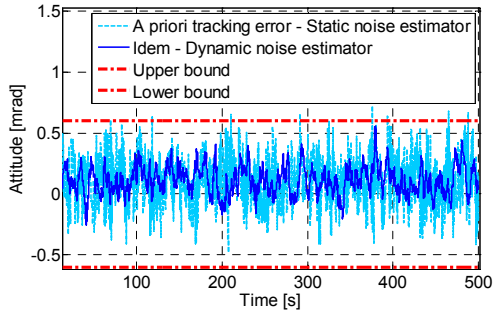

Fig. 8. A priori tracking error: dynamic and static noise estimator.

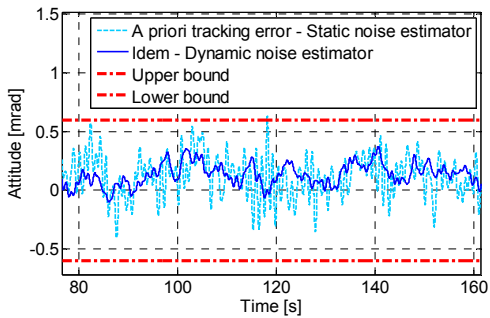

Fig. 9. Enlargement of Fig. 8.

### 3.3 Uncertainty-based control design

#### 3.3.1 Basic concepts

The uncertainty-based design aims to bound the a posteriori model error $\overline{\mathbf{e}}$ in (25) and (26). Equation (25) can be rewritten as

$$\overline{\mathbf{e}}(i) = C_c \hat{\mathbf{\eta}}_c(i) + \left( \mathbf{e}(i) + \mathbf{d}_y(i) \right), \quad (40)$$

where $\mathbf{d}_y$ has been defined in (31) and $\hat{\mathbf{\eta}}_c$, the corrected prediction error defined in (32), satisfies the error equation

$$\begin{aligned}
\hat{\mathbf{\eta}}_c(i+1) &= (A_c - G_c L C_c) \hat{\mathbf{\eta}}_c(i) - H_c \hat{\mathbf{x}}_d(i) - \\
&\quad - G_c N \mathbf{q}(i) - G_c L \left( \mathbf{e}(i) + \mathbf{d}_y(i) \right) \\
\hat{\mathbf{x}}_d(i+1) &= G_d L C_c \hat{\mathbf{\eta}}_c(i) + A_d \hat{\mathbf{x}}_d(i) + \\
&\quad + G_d N \mathbf{q}(i) + G_d L \left( \mathbf{e}(i) + \mathbf{d}_y(i) \right) \\
\mathbf{q}(i+1) &= L_q C_c \hat{\mathbf{\eta}}_c(i) + A_q \mathbf{q}(i) + \\
&\quad + L_q \left( \mathbf{e}(i) + \mathbf{d}_y(i) \right)
\end{aligned} \quad (41)$$

The forced response of (41) and (40) can be written in terms of the state predictor sensitivity $\mathbf{S}_m$ and of the complement $\mathbf{V}_m$ as follows

$$\begin{aligned}
\overline{\mathbf{e}} &= (I - \mathbf{V}_m) \cdot (\mathbf{e} + \mathbf{d}_y) = \mathbf{S}_m \cdot (\mathbf{e} + \mathbf{d}_y) \\
\mathbf{V}_m \cdot (\mathbf{e} + \mathbf{d}_y)(i) &= \sum_{k=1}^{i} C_m A_m^{i-k} B_m (\mathbf{e} + \mathbf{d}_y)(k-1)
\end{aligned}, \quad (42)$$

where $C_m$, $A_m$ and $B_m$ are the state predictor matrices obtained from (41).

Under Kalman assumption $\mathbf{e} = \mathbf{w}_y$ (the measurement noise) and $\mathbf{d}_y$ only depends on $\mathbf{w}$. Therefore, the covariance of $\bar{\mathbf{e}}$ is bounded and minimized by an optimal design of $L$, $N$, $L_q$ and $A_q$ in (41).

Abandoning Kalman assumption opens the 'robust' control domain: the goal is still to bound $\bar{\mathbf{e}}$, but in the presence of a neglected dynamics (15) and cross-coupling errors (37). The tool is the 'error loop' already mentioned in [2] and illustrated in Fig. 10. The loop shows how the state predictor filters the plant-to-model uncertainty inside a closed-loop starting from the model output $\mathbf{y}_m$ on the left of Fig. 10 and ending into the a priori tracking error $\mathbf{e}$ on the right. The loop summing point satisfies the equation

$$\mathbf{y}_m(i) = C_c(\underline{\mathbf{x}} - \underline{\mathbf{e}}_c)(i) = C_c(\mathbf{x}_c + Q\mathbf{x}_d) = C_c\mathbf{x}_c, \quad (43)$$

because of the Sylvester equation (22).

Uncertainty enters the loop from three channels:
1) the input signals $\mathbf{w}$ and $\mathbf{w}_y$ (exogenous uncertainty)
2) the fractional error dynamics $\mathbf{E}$, driven by $\mathbf{y}_m$, which is part of the loop (internal uncertainty),
3) the cross-coupling error $\Delta\mathbf{m}(\cdot)$, driven by $\mathbf{x}_c$, which is a component of $\mathbf{d} - B_c\underline{\mathbf{m}}(\cdot)$ (internal uncertainty).

A further input signal is $\underline{\mathbf{x}}$ and the output signal is $\underline{\mathbf{e}}$. As shown in Fig. 10, Kalman assumption opens the error loop. In this case, stability and performance are guaranteed by Theorem 1 and the design of $\mathbf{S}_m$.

Assume for simplicity's sake that $\mathbf{m}(\cdot)$ can be rewritten in terms of $\mathbf{y}_m$ and $\underline{\mathbf{m}}(\cdot)$ in terms of $\hat{\mathbf{y}}_m$; which is possible though a suitable dynamic feedback around the controllable dynamics (3). Then assume the anti-causal limit (38) holds, and consider the following first-order, exact developments of $\mathbf{E}(\cdot)$ and $\Delta\mathbf{m}(\cdot)$, given by

$$\mathbf{E}(\mathbf{y}_m) = \mathbf{E}(\tilde{\mathbf{y}}) - \partial\mathbf{E}(\tilde{\mathbf{y}}) \cdot \underline{\mathbf{e}}_y, \quad (44)$$
$$\tilde{\mathbf{y}} = \underline{\mathbf{y}} - \alpha\underline{\mathbf{e}}_y, \; 0 \le \alpha \le 1$$

and

$$\mathbf{m}(\mathbf{y}_m) - \underline{\mathbf{m}}(\hat{\mathbf{y}}_m) = \Delta\mathbf{m}(\tilde{\mathbf{y}}) - \partial\mathbf{m}(\tilde{\mathbf{y}}) \cdot \underline{\mathbf{e}}_y. \quad (45)$$

The following theorem provides the forced response of $\underline{\mathbf{e}}_y$.

*Theorem 2. Error-loop equation.* The forced response of the a priori tracking error is found to be

$$\begin{aligned}\left(I + \mathbf{V}_m \cdot \partial\mathbf{E}(\tilde{\mathbf{y}}) - \mathbf{S}_m \cdot \mathbf{M} \cdot \partial\mathbf{m}(\tilde{\mathbf{y}})\right) \cdot \underline{\mathbf{e}}_y = \\ = \mathbf{V}_m \cdot \left(\mathbf{E}(\tilde{\mathbf{y}}) + \mathbf{w}_y\right) - \mathbf{S}_m \cdot \mathbf{M} \cdot \left(\Delta\mathbf{m}(\tilde{\mathbf{y}}) + \mathbf{D} \cdot \mathbf{w}\right)\end{aligned}. \quad (46)$$

*Proof.* With the help of (40) and (42), of the anti-causal limit (38), of Fig. 10 and of the equalities

$$\underline{\mathbf{e}}_y = C_c\underline{\mathbf{e}} = C_c(\boldsymbol{\eta} - \mathbf{x}_y) = -\hat{\underline{\mathbf{e}}}_y = -\bar{\mathbf{e}} + \mathbf{e}, \quad (47)$$

the forced response of the a priori tracking error can be expressed in terms of output variables and errors as follows

$$\underline{\mathbf{e}}_y = (I - \mathbf{S}_m)\mathbf{e} - \mathbf{S}_m \cdot \mathbf{d}_y. \quad (48)$$

Separating exogenous and internal uncertainty in (48), one obtains

$$\underline{\mathbf{e}}_y = \mathbf{V}_m \cdot \left(\mathbf{E}(\mathbf{y}_m) + \mathbf{w}_y\right) - \mathbf{S}_m \cdot \mathbf{M} \cdot \left(\Delta\mathbf{m}(\cdot) + \mathbf{D} \cdot \mathbf{w}\right), (49)$$

where the output disturbance $\mathbf{d}_y$ has been split into the cross-coupling error $\Delta\mathbf{m}(\cdot)$ and stochastic disturbance $\mathbf{D} \cdot \mathbf{w}$, and the operators $\mathbf{M}$ and $\mathbf{D}$ derive from (3) and (9) respectively. Replacing $\mathbf{E}(\mathbf{y}_m)$ and $\Delta\mathbf{m}(\cdot)$ with their first order expansions (44) and (45), and shifting all terms driven by $\underline{\mathbf{e}}_y$ to left-hand side, (46) is proven.

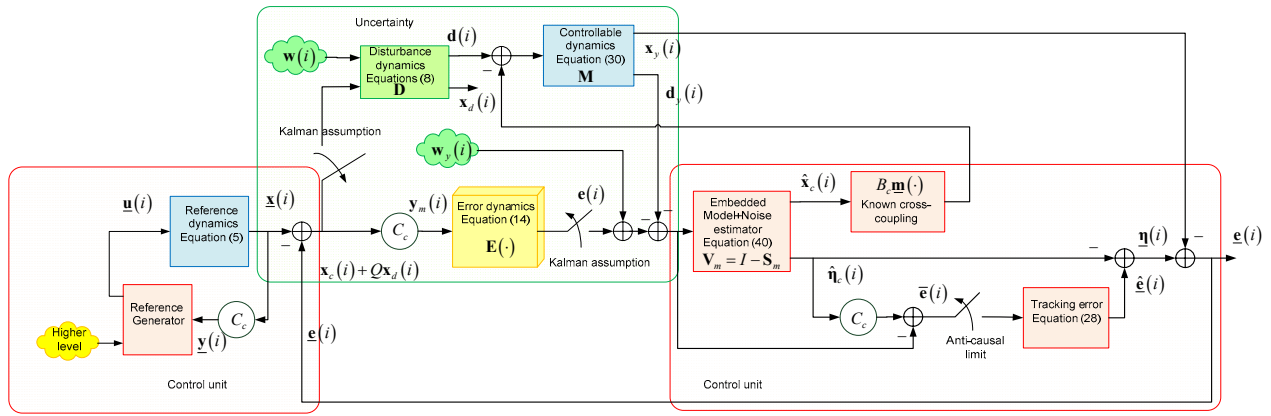

Fig. 10. Block-diagram of the error loop.

The error-loop equation (46) is the foundation of the uncertainty-based design. To this end, all uncertainty terms in (46) are assumed to be either bounded signals as $\mathbf{w}$, $\mathbf{w}_y$ $\mathbf{E}(\cdot)$ and $\Delta\mathbf{m}(.)$, or input-output bounded operators as $\partial\mathbf{E}(\cdot)$ and $\partial\mathbf{m}(\cdot)$. Operator uncertainty may be formulated through bounded sets $\partial\mathcal{E}(\cdot)$ and $\partial m(\cdot)$ where input-output norms $\|\partial\mathbf{E}(\cdot)\|$ and $\|\partial\mathbf{m}(\cdot)\|$ can be maximized. The assumption corresponds to finding a bounded envelope of the uncertainty as in the standard robust control design.

Parametrization of the operators allows to express the max norm as

$$\begin{aligned}\partial E_{\max} = \max_{\mathbf{p}\in\mathcal{P}, \mathbf{y}\in\mathcal{Y}} \left\|\partial\mathbf{E}(\mathbf{y};\mathbf{p})\right\| \\ \partial m_{\max} = \max_{\mathbf{n}\in\mathcal{N}, \mathbf{y}\in\mathcal{Y}} \left\|\partial\mathbf{E}(\mathbf{y};\mathbf{n})\right\|\end{aligned}, \quad (50)$$

where $\mathcal{Y}$ is the admissible reference set, and $\alpha = 0$ has been assumed in (44). Parametrization is essential in Monte Carlo runs, where parameters are extracted to cover the whole parameter set.

Guaranteeing $\overline{\mathbf{e}}$ to be bounded, which is the design aim, passes through the following Lemma.

*Lemma 1.* The a posteriori model error $\overline{\mathbf{e}}$ is bounded if and only if the a priori tracking error $\underline{\mathbf{e}}_y$ is bounded. The latter is bounded if and only if the right-hand side of (46) is bounded and the sensitivity operator

$$\Delta \mathbf{S} = (I + \Delta \mathbf{G})^{-1} = (I + \mathbf{V}_m \cdot \partial \mathbf{E}(\cdot) - \mathbf{S}_m \cdot \mathbf{M} \cdot \partial \mathbf{m}(\cdot))^{-1} \quad (51)$$

is input-output bounded. $\Delta \mathbf{G}$ is the error-loop operator.

*Proof.* The first part follows from (47) and from (42), since the latter ensures no cancellation of unbounded components occurs. The second part follows from solving (46) in terms of $\underline{\mathbf{e}}_y$.

The small-gain theorem [21] provides a sufficient condition for $\Delta \mathbf{S}$ in (51) to be input-output bounded.

*Theorem 3.* A sufficient condition for $\Delta \mathbf{S}$ to be input-output bounded is that

$$\max_{\partial \mathbf{E} \in \partial \mathcal{E},\ \partial \mathbf{m} \in \partial m,\ \mathbf{y} \in \mathcal{Y}} \|\Delta \mathbf{G}\| \leq \eta < 1, \quad (52)$$

where $\|\Delta \mathbf{G}\|$ is an input-output norm.

Assuming LTI worst-case approximations of $\partial \mathbf{E}(\cdot)$ and $\partial \mathbf{m}(\cdot)$ exist within $\mathcal{Y}$, (52) can be restated using the $H_\infty$ norm as follows:

$$\max_{\mathbf{p} \in \mathcal{P},\ \mathbf{n} \in \mathcal{N},\ |f| < f_{\max}} |\Delta \mathbf{G}(f;\mathbf{p},\mathbf{n})| \leq$$
$$\leq \max_{\mathbf{p} \in \mathcal{P},\ |f| < f_{\max}} |\mathbf{V}_m \partial \mathbf{E}(\mathbf{p})(f)| + \quad , \quad (53)$$
$$+ \max_{\mathbf{n} \in \mathcal{N},\ |f| < f_{\max}} |\mathbf{S}_m \mathbf{M} \partial \mathbf{m}(\mathbf{n})(f)| \leq \eta < 1$$

where notations have been unchanged with respect to (51). Inequality (53) suggests the guidelines of the uncertainty-based design.

*Statement 11.* Under the 'anti-causal' limit (38), the instrument for reducing the uncertainty effects on $\overline{\mathbf{e}}$, so as to guarantee plant closed-loop stability and performance (uncertainty-based design), is the noise estimator which is part of the state predictor sensitivity $\mathbf{S}_m$. According to (53) two contrasting objectives must be met:

1) the neglected dynamics $\partial \mathbf{E}$ must be filtered by the sensitivity complement $\mathbf{V}_m$, a low-pass filter, in the same manner as the measurement noise $\mathbf{w}_y$ in (46),
2) the cross-coupling error $\partial \mathbf{m}$ (including parametric uncertainty) must be attenuated by the sensitivity $\mathbf{S}_m$, a high-pass filter, in the same manner as the stochastic disturbance $\mathbf{D} \cdot \mathbf{w}$ in (46).

In general a trade-off may be unfeasible, which fact requires a careful design of the disturbance dynamics (9), including state variables, noise and the known component of $\mathbf{m}(\cdot)$ in the way of recovering feasibility. The feedback gain $K$ may contribute to stability only abandoning the anti-causal limit, which may be urged by limited control authority. In the latter case $\mathbf{S}_m$ is replaced by the sensitivity of the overall control unit, which depends on the feedback gain $K$. Such an extension is not pursued here; see [2].

Exploiting low- and high-frequency asymptotes of $\mathbf{S}_m$ and $\mathbf{V}_m$, respectively, analytic inequalities can be derived from (53) and employed for a first-trial design as in [22], [23] and [24].

*3.3.2 Case study*

Fig. 11 shows the Nyquist plot of the scalar transfer function $\mathbf{E}(s)$ of the neglected dynamics (worst-case) in (20), before and after being passed through $\mathbf{V}_m(s)$. Because of the linearity of $\mathbf{E}(s)$, (44) simplifies to

$$\mathbf{E}(s)\mathbf{y}_m(s) = \mathbf{E}(s)(\underline{\mathbf{y}} - \underline{\mathbf{e}}_y)(s), \quad (54)$$

which implies $\partial \mathbf{E}(s) = \mathbf{E}(s)$. Since the low-frequency part of (20) cannot be attenuated by $\mathbf{V}_m(s)$, it can be converted into a cross-coupling term that can be found to be

$$\mathbf{M}(s)\partial \mathbf{m}(s) \cong -\frac{1}{s\tau} - \partial J. \quad (55)$$

The transfer function (55) having a single pole in the origin, can be attenuated well below the unit by a sensitivity having four zeros in the origin as the $\mathbf{S}_m$ resulting from (7), (10) and (12), and a bandwidth larger than $p(2\pi) < 5$ mHz as in Table 3.

Restricting to the first part of the inequality (53), a sufficient stability condition is

$$\max_{|f| \leq f_{\max}} |\mathbf{V}_m(f)\mathbf{E}(f)| < 1, \quad (56)$$

which is actually not met by either design, since (53) is just a sufficient condition. But the dynamic design is largely far from encircling $(-1,0)$ unlike the static case.

A more complete survey can be made by plotting the experimental RMS of the attitude and of the angular rate a priori tracking errors, of the command $a_u$ in (39) and of the torque transmitted by the flexible link (in acceleration units) versus the complementary eigenvalue $\gamma_q = 1 - \lambda_q = (2\pi T_q)^{-1} f_q$ of the state predictors resulting from the noise estimators (12) and (13). The frequency $f_q$ can be shown to be approach the state predictor bandwidth. Eigenvalues, assumed to be each other equal, uniquely fix the noise estimator gains.

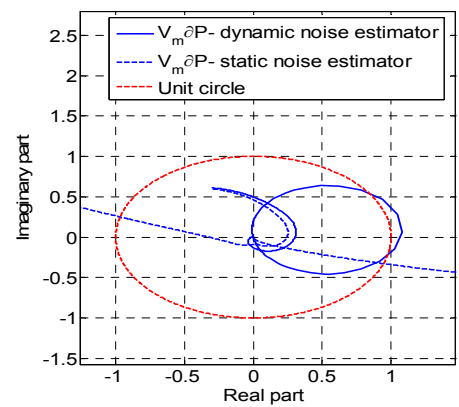

Fig. 11. Nyquist diagram of the fractional error dynamics.

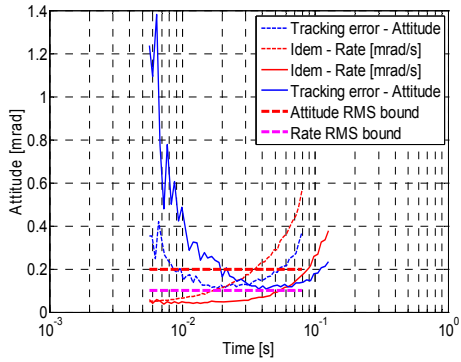

Fig. 12.  A priori tracking errors (rate and attitude) versus $\gamma_a$.

Fig. 12 plots the error RMS versus $\gamma_a$, thus indicating how much the bandwidth of the static case must be narrowed for repeating the dynamic case performance, typical of the EMC. All the simulated results refer to $\gamma_a = 0.03$, that is to $f_a \cong 0.05$ Hz, where static and dynamic performances are each other close.

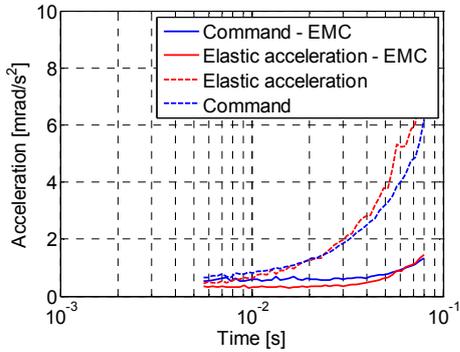

Fig. 13.  Commanded and transmitted acceleration.

The smaller stability margin of the static case for the same bandwidth, already pointed out in Fig. 11, can be better appreciated by comparing the command RMS (or which is the same the torque transmitted by the flexible link) in Fig. 13, under zero reference, i.e. for $t \geq 300$ s in Fig. 14. Above $\gamma_a = 0.01$ the RMS of the static case (dashed line) increases approaching instability, whereas the dynamic design (solid line) keeps the same effort over a large band, thus revealing robustness.

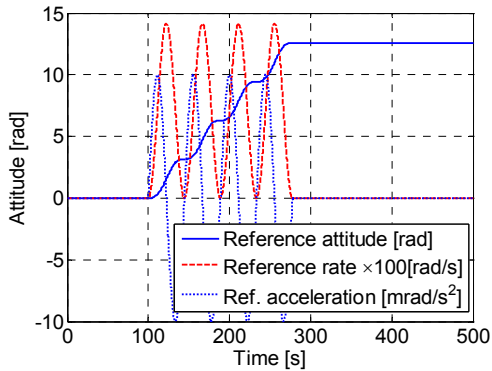

Fig. 14.  Reference signals.

All simulated results assume the reference signals in Fig. 14, that aim to progressively rotate the payload on the top of the flexible link by half a turn wide steps, while respecting the acceleration slew rate in Section 2.2.2.

Fig. 15 shows the command acceleration for both noise estimator design. The static design approaches the command bound of 0.025 rad/s². Fig. 16 compares the simulated disturbance (solid line) including drag, gravity, and other torques, with the rejected disturbance (dashed line) estimated by the embedded model. It is evident that the rejected disturbance is capable of implicitly estimating the neglected friction (time constant $\tau$) and the parametric uncertainty $\partial J$ of the body inertia, as they are amplified by the reference command in Fig. 14 during the interval $100 \leq t \leq 300$ s. The friction component being proportional to the angular rate is the large, negative triangular wave in Fig. 16, whereas $\partial J$ gives rise to a zero mean wave of smaller amplitude, proportional to acceleration in Fig. 14. In this manner, the model-based control law (39) is fully relieved from being designed against uncertainties.

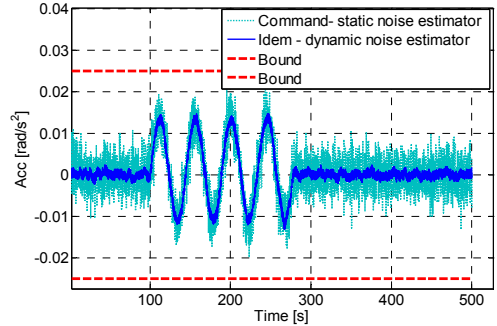

Fig. 15.  Commanded acceleration.

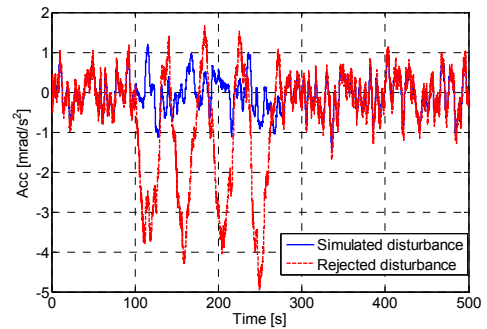

Fig. 16.  Simulated and rejected disturbance.

As a conclusion, the list of the complementary eigenvalues $\gamma_k = 1 - \lambda_k$, $k = 1,...,n$ of the different control sections is reported in Table 2.

Table 2 Complementary eigenvalues

| No | Control section | Size $n$ | Value $\gamma_k$ | Frequency $f_k$ [Hz] | Nyquist $f_{max}$ [Hz] |
|---|---|---|---|---|---|
| 0 | State predictor – attitude (12) | 3 | 0.03 | 0.048 | 5 |
| 1 | Idem –rate (12) | 3 | 0.03 | 0.48 | 50 |
| 2 | Feedback control (39) | 2 | 0.1 | 1.6 | 50 |

As anticipated in Section 3.2, the feedback control eigenvalues fixing the gains $k_q$ and $k_\omega$ in (39) are the fastest. They have been fixed somewhat below the Nyquist frequency $f_{max}$ to guarantee the command effort to stay below the bound with some margin as shown in Fig. 15.

## 4 CONCLUSIONS

As a key result the model-based control theorem affirms a control law under the anti-causal limit can be designed on the basis of the embedded model, which includes a disturbance dynamics to account for plant-to-model uncertainties, but without reference to the unmodelled uncertainty. The latter corresponds to parametric and structural uncertainty, which is set to zero in the embedded model, but is available in the form of bounded sets in the simulated design model. This procedure allows to design simple, yet effective control algorithms. The unmodelled uncertainty must be accommodated by the noise estimator in charge of real-time estimating the noise, which is the unique feedback channel through which plant-to-model discrepancies are conveyed into the embedded model, where they are saved as disturbance state variables. The noise estimator is in charge of blocking the unmodelled uncertainty capable of destabilizing the plant from entering the embedded model. The blocked uncertainty constitutes the residual model error (plant minus model output) equal to the opposite of the whole control error. The key tool for the uncertainty-based design is the state predictor sensitivity which is tuneable through the relevant eigenvalues.

## 5 ACKNOWLEDGEMENTS